\documentclass[]{spie}  

\usepackage{amsmath,amsfonts,amssymb}
\usepackage{graphicx}
\usepackage[colorlinks=true, allcolors=blue]{hyperref}
\usepackage{multirow,bigdelim}

\def\CN2{\mbox{$C_N^2 \ $}}

\def\CT2{\mbox{$C_T^2 \ $}}

\def\sigmal2{\mbox{$\sigma ^{2}_{I} \ $}}

\title{Evaluation of filtering techniques to increase the reliability of meteo forecasts for ground-based telescopes} 

\author[a]{Alessio Turchi}
\author[a,b]{Gianluca Martelloni}
\author[a]{Elena Masciadri}
\affil[a]{INAF - Osservatorio Astrofisico di Arcetri, L.go E. Fermi 5, 50125  Florence, Italy}
\affil[b]{INSTM - National Interuniversity Consortium of Materials Science and Technology, Via della Lastruccia 3-13, 50019 Sesto Fiorentino (Florence), Italy}

\authorinfo{Correspondence to Alessio Turchi: aturchi@arcetri.inaf.it}

\begin{document} 
\maketitle 

\begin{abstract}
In this contribution we evaluate the impact of filtering techniques in enhancing the accuracy of forecasts of optical turbulence and atmospheric parameters critical for ground-based telescopes. These techniques make use of the data continuously provided by the telescope sensors and instruments to improve the performances of real-time forecasts which have an impact on the telescope operation. In previous works we have already shown how a mesoscale high-frequency forecast (Meso-NH [\cite{lafore98,LAC2018}] and Astro-Meso-Nh models [\cite{masciadri99a,masciadri2017}]) can produce reliable predictions of different atmospheric parameters [\cite{turchi2017}] and the optical turbulence [\cite{masciadri2017}]. The mesoscale forecast has an advantage on the global model in having a better implementation of the physical atmospheric processes, including turbulence, and produces an output with greater spatial resolution (up to 100m or beyond). Filtering techniques that make use of the real-time sensor data at the telescope may help in removing potential biases and trends which have an impact on short term mesoscale forecast and, as a consequence, may increase the accuracy of the final output. Given the complexity and cost of present and future top-class telescope installations, each improvement of forecasts of future observing conditions will definitely help in better allocating observing time, especially in queue-mode operation, and will definitely benefit the scientific community in medium-long term.
\end{abstract}



\section{INTRODUCTION}
\label{sec:intro}
The daily operation of a top-class telescope installation requires more and more work with each step of increase of the overall size and complexity of the project. Given the large budgets allocated for the new ELTs that are being planned for the next decade, the cost of each missed night of observation, either due to technical or environmental problems, is becoming incredibly high. Also, the availability of complex adaptive optics (AO) facilities, which require specific atmospheric conditions in order to achieve optimal performances, demands an even more careful planning of the observation schedule, i.e. service mode. Thus, the availability of accurate weather forecasts that include either the atmospheric parameters (temperature, wind speed and direction, relative humidity, precipitable water vapor) and the most common astroclimatic parameters (seeing $\epsilon_0$, wavefront coherence time $\tau_0$, isoplanatic angle $\theta_0$) gives to the telescope operator a fundamental tool to optimally tune the telescope and AO operations. In Masciadri et al. [\cite{masciadri2013}] the scientific challenges related to the forecast of optical turbulence (OT) and atmospheric parameters at ground are explained in detail. Previous feasibility studies on Chilean sites such as VLT [\cite{masciadri2017,lascaux2015,masciadri2013,lascaux2013}], have shown how this tool can give crucial and reliable information on multiple fundamental parameters. The ALTA system (\footnote{\url{http://alta.arcetri.astro.it}}) is already operational at LBT and is providing validated forecast for all the above mentioned parameters on a daily basis. The system has already proven its reliability [\cite{turchi2017,adoni2018}] and is a key part of the new LBT observational strategy to maximize telescope scientific output [\cite{veillet2016}]. Model predictions, in a practical operational setup, are typically made available the day before the observing night, however there is also room for improvement on short-term reliability of the forecast which still may be useful to support night operations in real-time.\\
Despite the increased reliability of atmospherical modeling itself, the information given from the many atmospheric sensors present on telescope installations is presently wasted and not really used in our simulations (except for validation studies). The basic idea behind the preliminary analysis explained in this contribution, is that data assimilation and filtering techniques that make use of the real-time sensor data at the telescope may help in removing potential biases and trends which have an impact on short term mesoscale forecast and may also increase the accuracy of the final output by a noticeable margin.\\
From general considerations we do expect that adding more information from the real-time sensors on the site under study, despite being a spatially localized information, may help in reducing the error on short timescales on that specific point.
We still expect that on long timescales the model output will be better or equal to any forecast obtained through such ``local'' corrections. However we argue that each small increase in the forecast accuracy may have an impact on the telescope operations, which may be able to switch the scientific program during the course of the night if favorable conditions show up. This is particularly true in the case of ``opportunistic'' scientific observations that require rare atmospheric conditions, e.g. in terms of extremely low seeing or total water content in the atmosphere [\cite{kerber2014}].\\
Many techniques can be employed to exploit the local availability of real-time measurements. In the present study we focus on autoregression methods [\cite{Dzhaparidze1994}] paired with a Kalman filter [\cite{Kalman1960}]. While there are many implementations of such technique, all of them need a continuous data stream of measurements from sensors and the time series forecasted by a model. With these two ingredients, the effect of the filter is to remove spurious noise and correct the model forecast by reducing the difference with the measurements.
In particular the Kalman filtering techniques are already used for atmospheric data assimilation [\cite{Zheng2010}].\\
It is our interest to evaluate the performance of neural network techniques, however this will be the subject of a future work.\\
In section \ref{sec:filter} we will describe the initial implementation of the filtering techniques that we considered in this analysis. In section \ref{sec:model} we will describe the setup configuration on the two chosen sites, the parameters chosen for the analysis and the selected sample. In section \ref{sec:res} we will present the results of the analysis in terms of statistical operators. Finally in section \ref{sec:concl} we will draw the conclusions.\\

\section{FILTER TECHNIQUES}
\label{sec:filter}
In this paper we propose a method to improve the forecasting of the model Astro-Meso-NH up to first 4 hours (nowcasting). Our technique is based on the evaluation of the model error compared to measurements, assumed as ``true'' values without any error. We assume the availability of a continuous time series of measured values $V(t)$ and model forecasts $H(t)$. We resample both measurements and model forecasts with one minute step, in order to have the same sampling time for both time series. We then define the error time series $e(t)$ at time $t$ as the difference between the measure and the correspondent model forecast $H(t)$. 
\begin{equation}
e(t)=V(t)-H(t),
\label{eq0}
\end{equation}
Then, for every time $t$, we then evaluate the error time series by means of autoregressive (AR) model [\cite{Dzhaparidze1994}] of order $n$. The AR model depends on coefficients $a_0$, $a_1$, ..., $a_n$:
\begin{equation}
e(t)=a_0 + a_1 e(t-1) + a_2 e(t-2) + ... + a_n e(t-n),
\label{eq1}
\end{equation}     
such constants are computed on a previous time interval (buffer) which length was varied from 24 to 72 hours. In section \ref{sec:res} we report the result for different length this buffer.\\
\begin{figure} [ht]
\begin{center}
\includegraphics[width=0.7\textwidth]{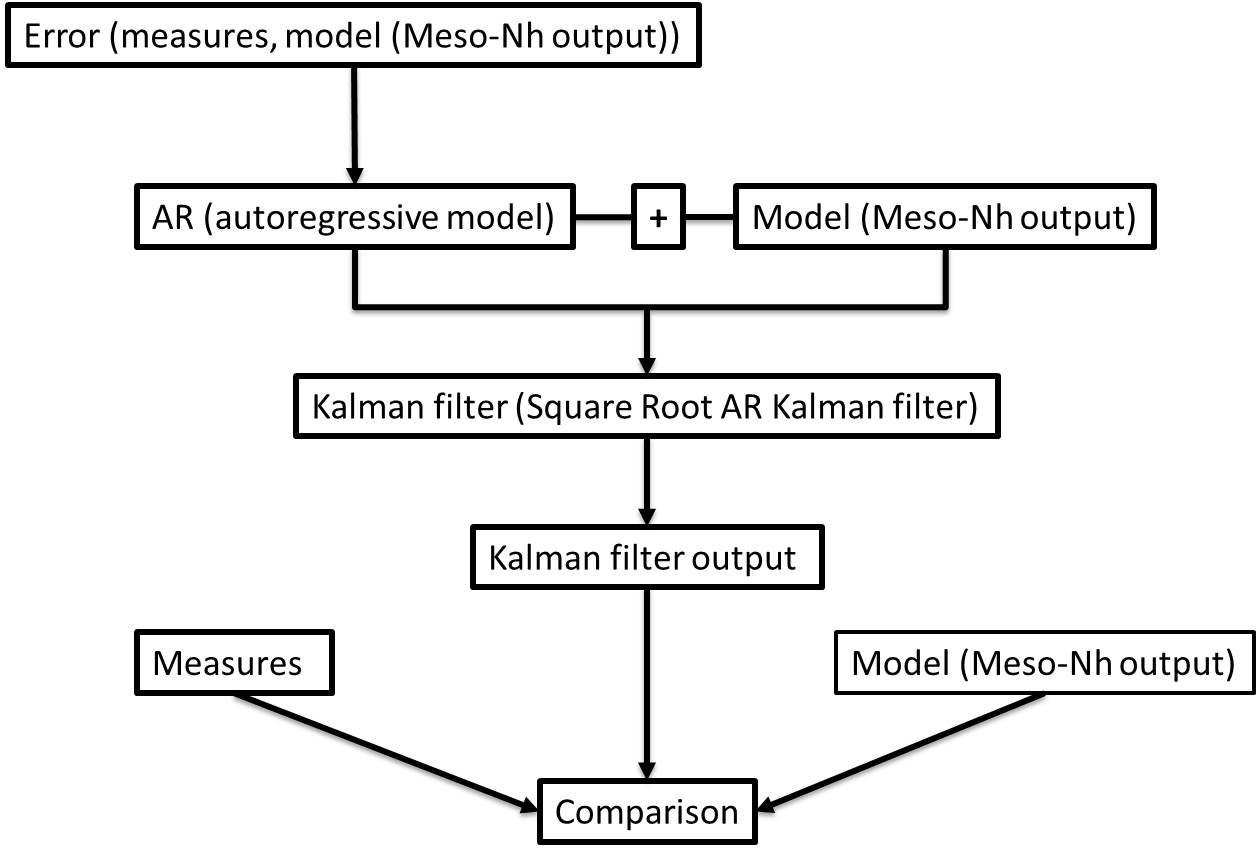} 
\end{center}
\caption{ \label{fig:kal} Operational scheme of the AR+Kalman filter procedure}
\end{figure} 
In this study we are interested in enhancing the forecasts of the nighttime period. While many atmospheric parameters have continuous measurements also during daytime, many fundamental parameters are measured only during the night (e.g. seeing $\epsilon_0$). For this reason we decided to discard every measurement relative to daytime an thus we have holes in our dataset corresponding to such periods. In principle this could reduce the efficacy of the AR procedure, however in order to overcome this limitation, we build our data buffer in order to have at least an hour worth of data after each hole. This allow the AR fit to ``jump'' the data hole without major loss of accuracy. Consequently the number of coefficients of the AR model is determined by testing the efficiency of the procedure in forecasting future data. Our preliminary results gives out reasonable results in term of accuracy and computational time. By using the AR function, for each time $t$, we can compute the forecasted errors on the required time (up to 4 hours). Moreover, summing the forecasted error with the corresponding model forecast, we obtain the corrected forecasted time series of the physical parameter under analysis. Finally we apply the Square Root Kalman Filter using U-D factorization to remove residual noise [\cite{Thornton1980}] in order to further enhance the forecast performance.\\ At present time we are also implementing a new algorithm based on neural network to possibly improve the results obtained with AR + Kalman Filter. While this latter algorithm is showing significant performance improvements, the heavy computational costs implied in such analysis will defer the study of the neural network results to a future work.

\section{TEST CONFIGURATION}
\label{sec:model}
In order to obtain a preliminary test of the reliability of filtering techniques in improving the model performances, we need the availability of continuous and reliable measurements in order to feed the autoregression algorithm. We selected two different test cases: the Large Binocular Telescope (LBT) at Mount Graham (Arizona) and the Very Large Telescope (VLT) at Cerro Paranal (Chile). In both cases there are continuous measurements available about all the relevant atmospheric parameters at ground. For this preliminary test configuration, concerning atmospheric parameters, we limit ourselves to test the efficacy of the filter with three different atmospheric parameters measured at the height of the telescope dome: temperature, relative humidity and wind speed. In the LBT case we also have the availability of all the forecast data produced by the ALTA project, which is performing operational forecasts for both atmospheric parameters and astroclimatic parameters (seeing $\epsilon_0$) on a daily basis, which are already validated [\cite{turchi2017}] and available for the telescope operations. However, since at LBT the DIMM measuring the seeing is placed inside the dome on the telescope frame, it doesn't guarantee the continuity of the measurements and it is only available when the telescope is observing. In this preliminary test we decided therefore to test the filter effects on the seeing on the VLT case, where we have the advantage of having continuous seeing measurements from the DIMM placed outside the dome. This is in addition to all the other atmospheric parameters which are also measured continuously by the weather stations.\\ 

For our simulations we use Meso-Nh\footnote{\url{http://mesonh.aero.obs-mip.fr/mesonh/}} [\cite{lafore98,LAC2018}] code, which is an atmospherical mesoscale model that simulates the time evolution of weather parameters in a three-dimensional volume over a finite geographical area. The coordinate system is based on mercator projection, since we are at low latitudes, while the vertical levels use the Gal-Chen and Sommerville coordinate systems (Gal-Chen et. al. 1975 [\cite{chen}]). We consider wave-radiation open boundary conditions with Sommerfeld equation for the normal velocity components (Carpenter et. al. 1982 [\cite{somm}]). The model is based on anelastic formulation of hydrodynamic equations, in order to filter acoustic waves. Simulations are made using a one-dimensional mixing length proposed by Bougeault and Lacarrere (Bougeault et. al. 1989 [\cite{Bougeault89}]) with a one-dimensional 1.5 closure scheme (Cuxart 2000 [\cite{Cuxart00}]). The exchange between surface and atmosphere is computed with the Interaction Soil Biosphere Atmosphere - ISBA scheme (Noilhan et. al. 1989 [\cite{Noilhan89}]).\\
The forecasts of astroclimatic parameters ($\epsilon_0$,$\tau_0$,$\theta_0$) are produced by the Astro-Meso-Nh code, developed by Masciadri et al. [\cite{masciadri99a}] and updated through the years [\cite{masciadri2017}]. For this specific kind of forecast, the Astro-Meso-Nh code needs to be calibrated with $C_N^2$ profiles measured on multiple nights [\cite{masciadri2001}]. At VLT we don't have mesoscale model forecasts available as an operational product. In contrast with LBT which has recently calibrated seeing forecasts, we only have an outdated calibration obtained for a different model version and different measurements and instruments relative to several years ago [\cite{masciadri2017}]. A new calibration is being produced for VLT with recent measurements but is still not yet completed and validated. In this preliminary analysis we are not interested in absolute performances, but only in the relative gain obtainable from the use of filtering techniques, so we decided to use the old calibration keeping in mind that the model performance will be worse for the seeing forecast. In this case it will be even more interesting to understand the filter contribution in presence of forecasts having expected sources of errors. Please note that the calibration has influence only on the astroclimatic parameter forecasts, while the atmospheric paramters don't need tobe calibrated for the specific site.\\

In the LBT test case we selected a total of 102 nights starting from 2016/09/21 to 2016/12/31. In the VLT case we selected a total of 109 nights starting from 2017/10/15 to 2018/01/31. In both cases we make use of initialization data obtained from the ECMWF general circulation. The model configuration used for LBT was already discussed in Turchi et al. 2017 [\cite{turchi2017}], while the configuration used for VLT was already presented in Lascaux et al. 2015 [\cite{lascaux2015}]. With respect to the works cited there, we must add that, in this work, the ECMWF initialization data have (since March 2016) an increased horizontal resolution of 9km (before was 16km), with a positive effect of the accuracy of mesoscale simulations.

In order to define the computation grid for the simulations, we use a grid-nesting technique (Stein et. al. 2000 [\cite{Stein00}]). This consists of using different imbricated domains, with digital elevation model (DEM, i.e. orography) extended on smaller and smaller surfaces having a progressively higher horizontal resolution. In this way, using the same vertical grid resolution, we can achieve a higher horizontal resolution on a sufficiently small scale around the summit to provide the best possible prediction at the specific site.\\
In the LBT test case most of the forecasts are obtained from the domain at 500m horizontal resolution, while the wind speed uses also the information coming from the 100m horizontal resolution domain as explained in [\cite{turchi2017}], which is fundamental for correctly resolving high wind speed (see Lascaux et al. 2013[\cite{lascaux2013}]). A similar configuration is used for the VLT case, which is explained in [\cite{lascaux2015}].\\
The model output have a high temporal sampling frequency ranging from few seconds (atmospheric parameters) to two minutes (astroclimatic parameters). This high frequency of the data was useful in correctly tuning the performance of the filtering techniques since it allows better stability an more accurate reconstruction of the AR process.\\

\section{PRELIMINARY TEST RESULTS WITH AR+KALMAN FILTER}
\label{sec:res}
In this section we report the tests that we performed on the initial implementation of the data assimilation and filter algorithms, as explained in section \ref{sec:filter}.\\
For this preliminary analysis we tested the effect of the filter on a short timescale. From any given point in time, we consider only the first 4 hours of the Kalman prediction and then update them each hour along the forecasted time. We thus have a 4-hours window moving in time that is updated each hour. When performing our statistical validation we consider all the time windows obtained with this method on all the nights of the sample and then proceed to compute the statistical indicators.

\begin{figure} [ht]
\begin{center}
\includegraphics[width=0.7\textwidth]{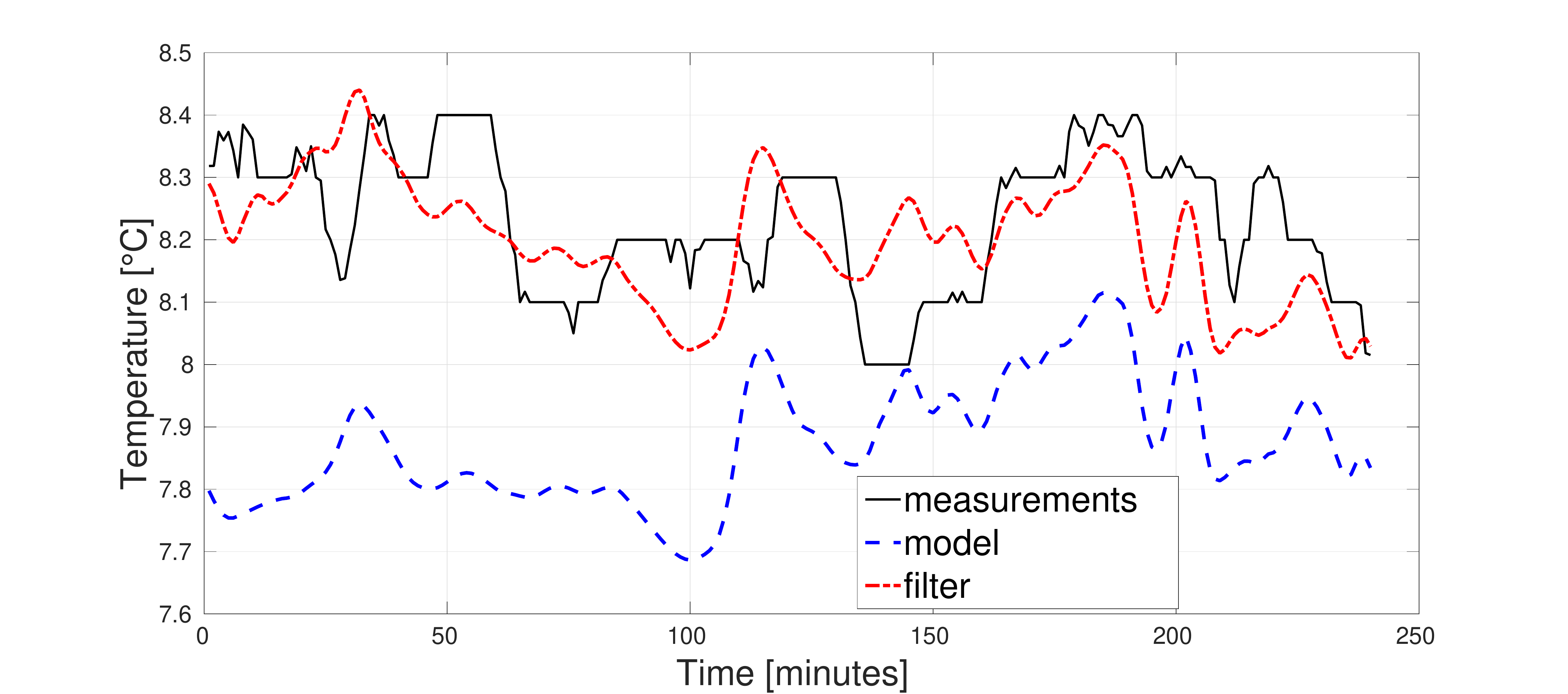} 
\end{center}
\caption{ \label{fig:ex} Example of the effect of the AR+Kalman filter on a 4-hour sample of the time evolution of temperature at LBT. The black full represents the temperature measured by the weather station, the blue line represents the prediction of the model, while the red line is the model prediction adjusted by the AR+Kalman filter.}
\end{figure}

In figure \ref{fig:ex} we report an example of the effect of the filter on a 4-hour time window of the evolution of the temperature parameter at ground measured on LBT. The model prediction is already quite accurate, being within a 0.5  $^{\circ}$C margin from the measure. The effect of the filter however is extremely effective since the agreement between the filtered prediction and measurements is almost perfect.\\

In order to quantify the system reliability in reconstructing the forecasted parameters we used the classical BIAS, RMSE and $\sigma$ statistical operators, defined as:
\begin{equation}
BIAS = \sum\limits_{i = 1}^N {\frac{{(Y_i  - X_i )^{} }}
{N}} 
\label{eq2}
\end{equation}
\begin{equation}RMSE = \sqrt {\sum\limits_{i = 1}^N {\frac{{(Y_i  - X_i )^2 }}
{N}} } 
\label{eq3}
\end{equation}
where $X_{i}$ are the individual observations and $Y_{i}$ the individual simulations (either filtered or directly taken from model output) calculated at the same time index $i$, with $1\leq i\leq N$, $N$ being the total sample size.\\
From the above quantities we deduce the bias-corected RMSE ($\sigma$):
\begin{equation}\sigma = \sqrt {RMSE^2 - BIAS^2}
\label{eq4}
\end{equation}
The previously defined indicators provide us information on the statistical and systematic errors.\\

\subsection{LBT INITIAL TEST}
In the LBT test case we tested three atmospheric paramteres at ground: temperature, relative humidity (RH) and wind speed. Each parameter is measured by the weather stations located on the telescope dome at $\sim$54m above the ground level. The wind measure is tricky because there are two anemometers placed on the front and on the rear of the dome, and each give a reliable measure only if the wind speed comes from a specific angle, as explained in Turchi et al. 2017 [\cite{turchi2017}].\\
As explained in section \ref{sec:filter}, the AR filter needs to have a buffer of past data in order to give an estimate about the future state of the system. In this analysis performed on LBT we considered only one past night worth of data. As already noted, we discard the daytime data, so our dataset have holes corresponding to such times. In order to overcome this we have to consider also the first hour of the simulated time, which typically falls in the twilight time and is not useful for the telescope operations. Thus in the LBT case, since simulated time starts at 00:00UT of each date [\cite{turchi2017}], we start to operate the filter from 01:00UT. This allows the AR procedure to corretly fit the jump in the dataset with almost no performance decrease.\\
\begin{figure} [h]
\begin{center}
\includegraphics[width=0.8\textwidth]{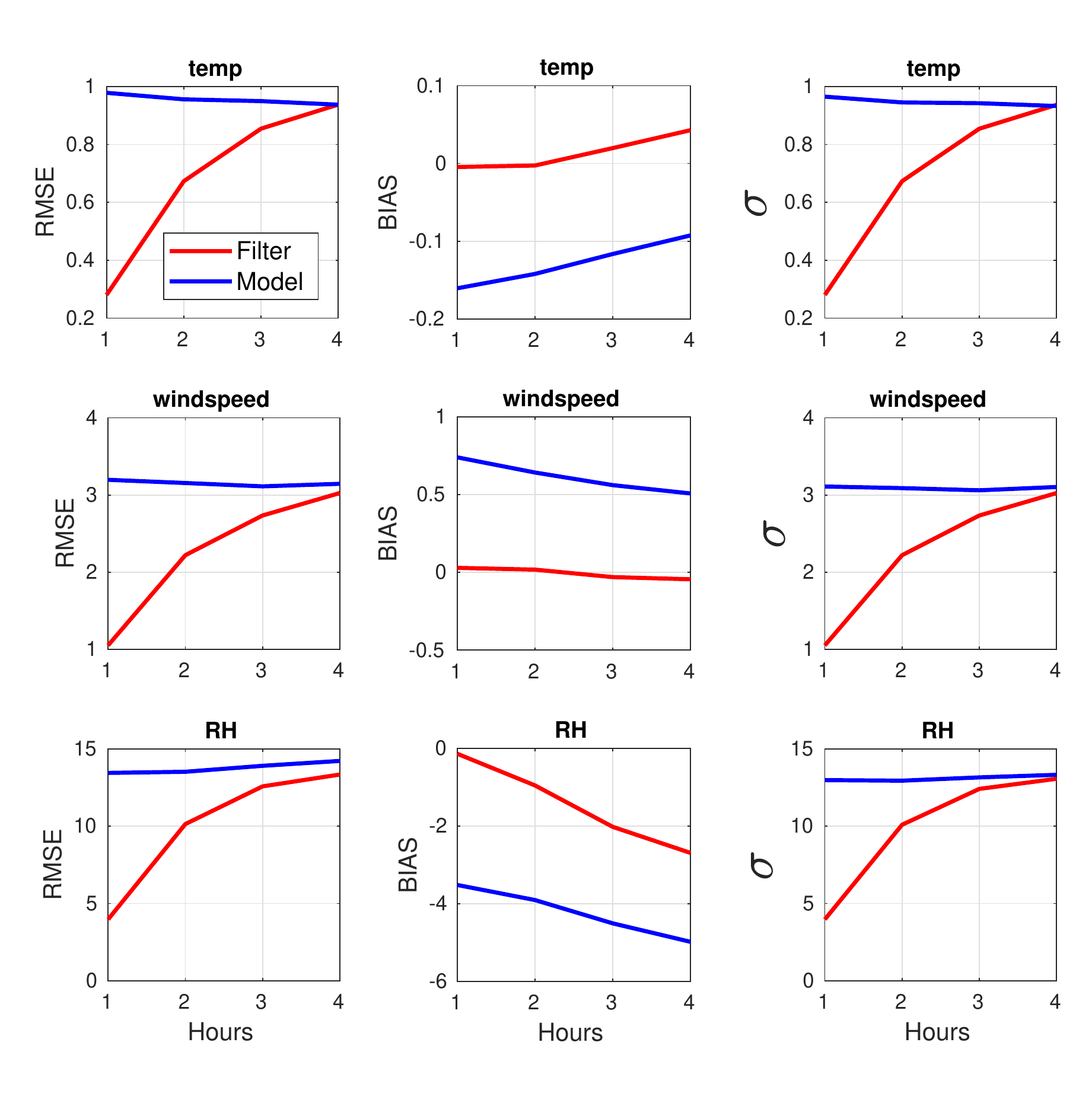} 
\end{center}
\caption{ \label{fig:mtg} 
LBT case: Performances of the filter technique expressed with the statistical operators RMSE, BIAS, σ, computed over the whole 102 nights sample on LBT for three parameters (temperature, wind speed, relative humidity). We report the performance over the four consecutive hours. Each point is computed cumulatively (e.g. the performance over 3 hours is the performance considering only the first 3 hours of the Kalman prediction). The blue line is the raw model predictions, while the red lines is the filtered predictions computed considering one nights of previous data. The filter performance is excellent in the first 1-3 hours and becomes similar to the model after the 4th hour in the future. This proves that the technique, albeit quite powerful, is useful only for short-timescale adjustments. After this time the physical (chaotic dynamics) sources of error are too large for the filter to compensate. Also the performance seems to saturate with a 2-days previous data buffer.}
\end{figure} 

In figure \ref{fig:mtg} we report the performance of the implemented method obtained with the above setup, considering all the 4 hours of the time window. Each point in the figure is computed cumulatively, i.e. the performance measured at the $N^{th}$ hour in the future corresponds to the statistical indicator computed on all the $N$ hours of the time window up to that point. We see that the filter has an excellent performance in the first 1-3 hours of the 4-hour window and is able to effectively reduce the model error by a large margin. In the temperature case, where the model has the best accuracy between the considered parameters (typically an error below 1 $^{\circ}$C, either considering RMSE or $\sigma$), the filter is extremely effective in the first hour, by reducing the model error by a factor $\sim$4, while the performance decrease reaching the same level of confidence as the raw model by the 4th hour in the future. In the wind speed and relative humidity cases the effect of the filter is still fundamental, however the error reduction factor is lower (respectivel $\sim$3 and $\sim$2.5 in the first hour). Still we see that around the 4th hour the performance has decayed and there is no clear advantage of the filtered result over the model prediction.\\
In all the above cases wee see that the filter is most effective at reducing the bias term of the error. Specifically the performance on the bias doesn't visibly decay even after the 4th hour. Considering the RMSE, over a large enough time the random components of the error, mainly due to the chaotic dynamic typical of an atmospheric system and thus scarcely predictable by definition, are dominant with respect to the bias and thus the potential gain is reduced almost to zero.\\

\subsection{VLT TEST}
In the VLT test case we replicated the test performed for LBT. In this case we also tested how the length of the previous data buffer, over which we train the AR filter, impacts on the global performance of the filtering process. We refer the reader to Lascaux et al 2015 [\cite{lascaux2015}] for a more specific description of the VLT sensor network. As As a first benchmark test we used only the temperature, wind speed and relative humidity sensors which are placed at 30m altitude above the groun of Cerro Paranal plateau. Also in this case we discard daytime, however VLT simulations start at 18:00UT [\cite{lascaux2015}], while the night time typically starts 5-6 hours later. Also in this case we select the first daytime/twilight portion of the data in order to overcome the holes in the dataset during the AR procedure.\\

\begin{figure} [h]
\begin{center}
\includegraphics[width=\textwidth]{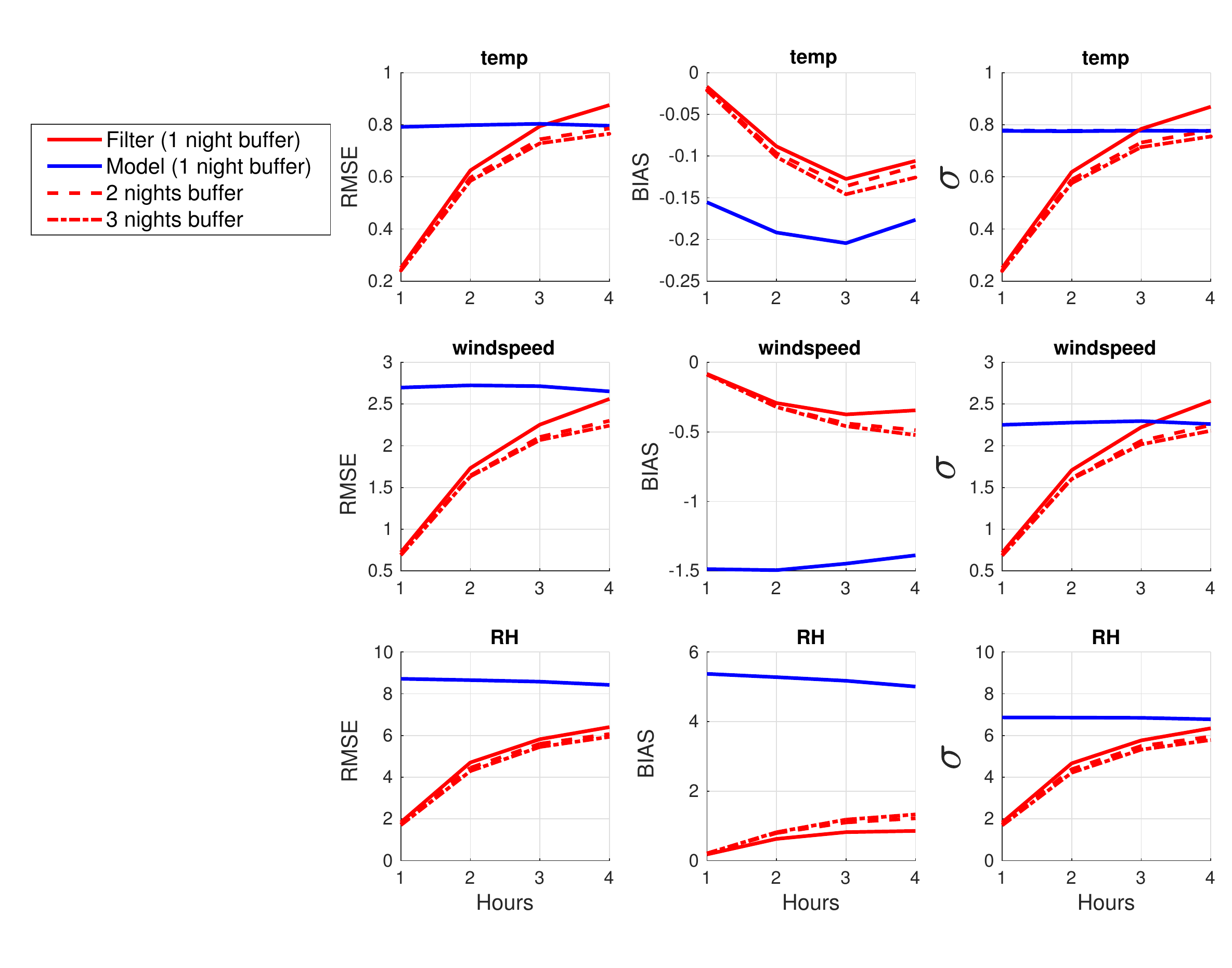} 
\end{center}
\caption{ \label{fig:par} 
VLT case: Performances of the filter technique expressed with the statistical operators RMSE, BIAS, σ, computed over the whole 109 nights sample on VLT for three parameters (temperature, wind speed, relative humidity). We report the performance over the four consecutive hours. Each point is computed cumulatively (e.g. the performance over 3 hours is the performance considering only the first 3 hours of the Kalman prediction). The blue line is the raw model predictions, while the different red lines are the filtered predictions computed considering a different buffer of previous data (1, 2 and 3 nights). As already seen in the LBT case the filter performs optimally in the first 1-3 hours, while after the 4th hour the benefit with respect to the model is neglectable. Also the performance seems to saturate with a 2-days previous data buffer.}
\end{figure}

In figure \ref{fig:par} we report the results of the test with one, two or three days lenght of the previous data buffer. As in the LBT case, we are evaluating cumulatively the 4-hours filter prediction window with the statistical indicators computed over all the possible 4-hours windows on the selected 109 nights sample. We see that the performace significantly increase by adding a second night of past data, however by adding a third night performance saturate, meaning that no further useful information is added to the AR procedure.\\
In the temperature case, which is the most accurate prediction of the model, we see that both RMSE and $\sigma$ can't be decreased after the 4th hour. In the wind speed and relative humidity cases, while $\sigma$ still is almost the same as the model after the 4th hour, we see that the filter continue in decreasing the RMSE by a noticeable margin at the end of the time window. This implies that by the 4th hour the RMSE still have important bias components over which a filtering technique can be very effective. In a future study it will be important to extend the length of the time window beyond the 4 hours.\\

\begin{figure} [h]
\begin{center}
\includegraphics[width=\textwidth]{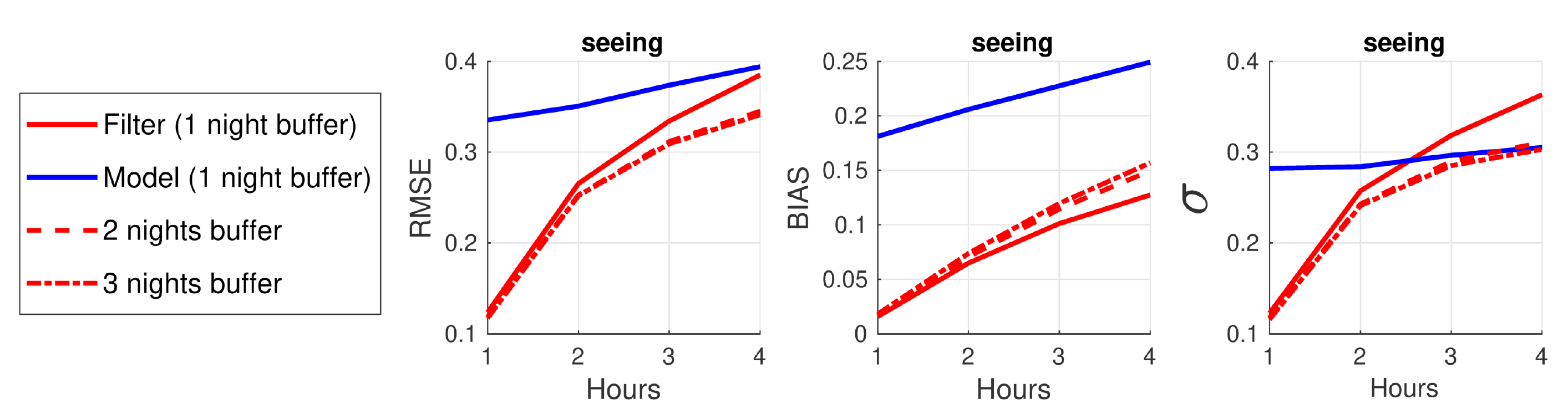} 
\end{center}
\caption{ \label{fig:parsee} 
Performances of the filter technique with respect to the seeing parameter, as measured from the DIMM at Paranal. Here we specify that the model output is not calibrated, as it should be for an accurate forecast of astroclimatic parameter. A study for an accurate calibration of CN2 profiles over VLT is ongoing, however here we are interested only in observing the potential gain coming from filtering techniques with respect to a generic reference. The gain from the Kalman filter extends over the 4-hours range (RMSE), however in terms of σ it still
saturates after 4 hours.}
\end{figure} 

We also tested the filter performances with the $\epsilon_0$ parameter, given the availability of DIMM measurements at VLT. We highlight the fact that, for the seeing, the model requires a calibration which make use of turbulence stratification measured in situ [\cite{masciadri2001}]. At present we are calibrating the model using the most recent measurements provided by the new instruments run by ESO at VLT: a new DIMM a and a Stereo-SCIDAR [\cite{osborn2018}]. It is however not necessary to have a calibrated and optimized version of the model to test the filter effects on the model forecasts. We decided to perform the test o the summer season, for which we have a model version calibrated with measurements related to a site testing campaign that took place some years ago (2007) [\cite{masciadri2017}]. This is enough to quantify if the filter has any effect on the seeing values forecasted by the Astro-Meso-NH model. Beside that, we note that typical values of the seeing are now very different from those obtained with the old DIMM and the Generalized SCIDAR used on Paranal site testing campaign of 2007. We expect therefore that the new calibration will provide smaller RMSE with respect to the one reported in this contribution. As a consequence, we expect even a better final result produced by the filtering technique.\\

In figure \ref{fig:parsee} we see that on the first hour the filter is able to reduce the model error by a factor 3 in terms of RMSE. Similarly to what is observed for the other atmospheric parameters, the $\sigma$ value becomes identical to the one given by the model after the 4th hour, however the filter is still removing arount 15\% of RMSE by that time, meaning that the predictable error sources are still an important factor. We also note that in this case it is fundamental to use a 2-nights buffer of past data in order optimize filer performance.\\
Our results indicate an excellent performance of the filtered model (even if not yet properly calibrated), beeing able to reduce the RMSE by a factor 2-3 in the first couple of hours of prediction. This means that there could be huge improvement in the short-term observation planning capability of a telescope. A recent paper [\cite{osborn2018}] has compared the relative accuracy of different instruments for seeing measurements, showing that typically each tool can only agree on a measured $\epsilon_0$ within an RMSE of order 0.2 arcsec. As seen in figure \ref{fig:parsee}, the filtered forecast, within the first two hours of the prediction, can get around that threshold or even better, even if starting from an initial model forecast that isn't properly calibrated. Based on these initial findings we have reasons to expect that the same analysis performed with an optimally calibrated model will give even better results.

\section{CONCLUSIONS}
\label{sec:concl} 
This study presents a preliminary implementation of a data assimilation and filtering technique that make use of the real-time measurements produced by the telescpe sensors in order to improve the reliability of numerical forecasts on short-term timescale. Specifically we have shown how a simple technique is able to give a huge improve, in terms of reduction of bias, RMSE and $\sigma$, over a time window of 4 hours in the future. Specifically we can decrese the RMSE by a factor 2-3 on the first hour, while the performance decrease from the second hour and becomes a small contribution by the 4th hour. In some cases the gain of the filter can persist even beyond the 4th hour. Despite the limits of this approach, we argue that a real-time implementation of such techniques, applied to the Astro-Meso-NH model characterized by high frequency forecasts, may have a positive impact in the short-term planning of the observation schedule during the night. A perfect example for the preliminary implementation of this scheme would be LBT, where high frequency forecasts are already available operationally through the ALTA project. Future work will focus in refining the techniques shown in this contribution.

\acknowledgments  
The ALTA Center project is funded by the LBTO. Measurements of surface parameters are provided by the LBT telemetry and annexed instrumentation. The authors thanks Christian Veillet, Director of the Large Binocular Telescope, for his continued and valuable support given to this research activity. Authors also thanks the LBTO staff for their technical support and collaboration. In site measurements for VLT are provided by ESO archive (\footnote{\url{http://archive.eso.org/cms/eso-data/ambient-conditions.html}}). We thank the MOSE ESO Board (Pierre-Yves Madec, Marc Sarazin, Florian Kerber, Harald Kuntshner) and Julien Milli, member of the ESO science operation team at Paranal for the support given to this study.



\end{document}